\documentclass[
 reprint,
superscriptaddress,
 amsmath,amssymb,
 aps,
longbibliography
]{revtex4-2}

\usepackage{physics}

\usepackage{graphicx}
\usepackage{dcolumn}
\usepackage{bm}
\usepackage{xcolor}

\newcommand{\MYhref}[2]{\href{#1}{\color{UBcolor}{#2}}}

\usepackage{mathtools}
\usepackage{gensymb}

\setcitestyle{super}
\usepackage{caption}
\usepackage{subcaption}
\DeclareCaptionLabelSeparator{bar}{~\rule[-0.4ex]{0.2ex}{1em}~}
\DeclareCaptionLabelFormat{subfor}{\textbf{#2}}
\newcommand*\bfcaption[2]{\caption[#1]{\textbf{#1.}#2}}
\usepackage{xcolor}
\definecolor{UBcolor}{HTML}{007CC1}
\usepackage[colorlinks=true,pdfnewwindow=true,linkcolor=UBcolor,citecolor=UBcolor,urlcolor=UBcolor,breaklinks=true,linktocpage]{hyperref}
\usepackage[all]{hypcap}
\usepackage[nameinlink,capitalise]{cleveref}
\crefname{SI section}{SI Section}{SI Sections}
\Crefname{SI section}{SI Section}{SI Sections}

\newcommand{\edited}[1]{{\color{black} #1}} 

\begin{document}

\preprint{APS/123-QED}

\title{Lattice-dependent orientational order in active crystals}

\author{Till Welker}
\email{t.a.welker@sms.ed.ac.uk}
\affiliation{School of Physics and Astronomy, University of Edinburgh, Peter Guthrie Tait Road, Edinburgh EH9 3FD, United Kingdom}
\affiliation{Max Planck Institute for the Physics of Complex Systems, N\"{o}thnitzerst. 38, 01187 Dresden, Germany}

\author{Ricard Alert}
\email{ralert@pks.mpg.de}
\affiliation{Max Planck Institute for the Physics of Complex Systems, N\"{o}thnitzerst. 38, 01187 Dresden, Germany}
\affiliation{Center for Systems Biology Dresden, Pfotenhauerst. 108, 01307 Dresden, Germany}
\affiliation{Cluster of Excellence Physics of Life, TU Dresden, 01062 Dresden, Germany}

\begin{abstract}
Via mechanisms not accessible at equilibrium, self-propelled particles can form phases with positional order, such as crystals, and with orientational order, such as polar flocks. However, the interplay between these two types of order remains relatively unexplored. Here, we address this point by studying crystals of active particles that turn either towards or away from each other, which can be experimentally realised with phoretic or Janus colloids or with elastically-coupled walker robots. We show that, depending on how these interactions vary with interparticle distance, the particles align along directions determined by the underlying crystalline lattice. To explain the results, we map the orientational dynamics of the active crystal onto a lattice of spins that interact via (anti-)ferromagnetic alignment with each other plus nematic alignment with the lattice directions. Our findings indicate that orientational and positional order can be strongly coupled in active crystals, thus suggesting strategies to control orientational order by engineering the underlying crystalline lattice.
\end{abstract}

\maketitle

In active matter, microscopic constituents inject mechanical energy, thus driving the system out of equilibrium. As a result, active particles can self-organize in ways not accessible at equilibrium. In particular, the field has focused on how positional and orientational order can emerge \cite{Marchetti2013,Cates2015,Marchetti2016,Fodor2018,Bar2020,Chate2020,Baconnier2025}.

Orientational order, such as the polar order found in flocks, can arise from direct alignment interactions between the orientations of self-propelled particles, as originally demonstrated in the Vicsek model \cite{Vicsek1995}. More recent work showed that flocking can also emerge when active particles attract each other \cite{Caprini2023}, align their orientation with their velocity \cite{Szabo2006,Ferrante2013,Baconnier2025} or, alternatively, when particles turn away from one another \cite{Knezevic2022,Das2024,Subramaniam2025}.

Regarding positional order, self-propelled particles have been found to crystallise via either motility-induced phase separation \cite{Redner2013,Palacci2013,Buttinoni2013,Digregorio2018,vanderLinden2019,Omar2021}, attractive interactions \cite{Mognetti2013,Singh2016,Thutupalli2018,Mauleon-Amieva2020,Caprini2023,Kole2025}, or simply at densities approaching close packing \cite{Bialke2012,Weber2014,Briand2016,Briand2018,Shi2023,Zhang2025,Sakai2025}. Recent work also showed that, in confinement, self-propelled particles can form Wigner crystals that emerge through repulsive interactions, which keep the particles at a distance \cite{Smeets2016,Delfau2017,LeBlay2022,Das2024,Yang2024,Subramaniam2025}. Particles in active crystals were also found to orient and move collectively as a flock, thus displaying not just positional but also orientational order \cite{Gregoire2003,Menzel2013,Briand2018,Baconnier2022,Caprini2023,Hernandez-Lopez2024,Das2024,Subramaniam2025,Musacchio2025}. Beyond such flocking crystals, the interplay between positional and orientational order in active matter \edited{has been recently} explored \edited{in the XY model with vision-cone interactions \cite{Loos2023,Bandini2025,Popli2025,Dopierala2025}, in crystals of self-aligning walker robots \cite{Baconnier2022,Hernandez-Lopez2024}, and in crystallites of Quincke rollers \cite{Kole2025}.}

Here, we address this question by studying crystals of self-propelled particles that turn either towards or away from each other. These interactions, which emerge for example in metal-dielectric Janus colloids \cite{Zhang2021i,Das2024}, couple the polarity of one particle to the orientation of the bond with a neighboring one. Hence, such polarity-bond interactions produce a crosstalk between positional and orientational order. We show that, on a lattice, polarity-bond interactions yield either effective alignment or anti-alignment between particle polarities, like in the XY model. In addition, they also produce nematic alignment of the particle polarities and the lattice axes. We first study the interplay between these two effects for particles on a chain. We find that the particles can achieve either local ferro- or antiferromagnetic order, either along or perpendicular to the chain. We then consider a square lattice and find that the particles can orient locally along the lattice axes and/or form domains of polar order, depending on the distance dependence of the underlying interactions. \edited{On the triangular lattice, the polarity-bond interactions can be frustrated.} Overall, our findings show that, through polarity-bond interactions, the orientational order of active crystals can depend strongly on the lattice structure. Thus, our work suggests strategies to obtain desired states of orientational order in active crystals by engineering specific particle interactions and crystalline lattices.

\bigskip

\noindent\textbf{Active crystals with polarity-bond interactions}

\begin{figure}[tb!]
    \centering
  {\phantomsubcaption\label{Fig:lattice}}
  {\phantomsubcaption\label{Fig:torques}}
  {\phantomsubcaption\label{Fig:distance_dependence}}
    \includegraphics[width=\columnwidth]{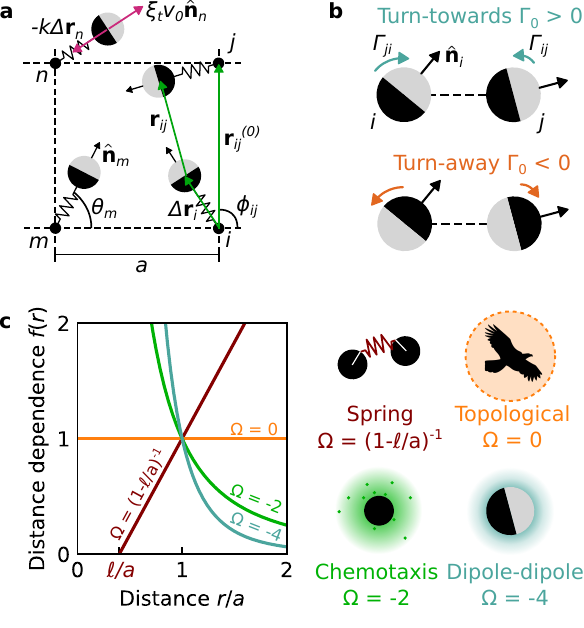}
    \bfcaption{Active crystal with polarity-bond interactions}{ \subref*{Fig:lattice}, Schematic of an active crystal made of self-propelled particles bound to lattice sites. The green arrows indicate distance vectors. The purple arrows indicate the self-propulsion force $\xi_\text{t} v_0 \hat{\mathbf{n}}_i$ and the elastic restoring force $-k \Delta \mathbf{r}_i$, whose balance sets the equilibrium displacement of a particle from its lattice site. \subref*{Fig:torques}, Polarity-bond interactions are torques, defined in \cref{eq:torque_begin}, whereby particles turn either towards or away from others. \subref*{Fig:distance_dependence}, Examples of distance dependences $f(r)$ of the interaction torques for different systems, with their corresponding dimensionless parameter $\Omega \equiv a f'(a)$.}
    \label{fig:trapped}
\end{figure}

We consider active particles on a fixed crystalline lattice. Neighboring lattice sites $i$ and $j$ are separated by the vector $\mathbf{r}^{(0)}_{ij} = a(\cos{\phi_{ij}},\sin{\phi_{ij}})$, where $a$ is the lattice constant and $\phi_{ij}$ define the lattice angles (\cref{Fig:lattice}). The particles are bound to lattice sites by elastic forces $-k\Delta\mathbf{r}_i$, with elastic constant $k$ and displacement $\Delta\mathbf{r}_i$ (\cref{Fig:lattice}). These elastic forces correspond to the harmonic approximation of any force that confines the particles to their lattice sites. In addition, the particles self-propel at speed $v_0$ along their orientation $\hat{\mathbf{n}}_i = (\cos\theta_i,\sin\theta_i)$.

The particles interact through turn-towards or turn-away torques given by
\begin{equation} \label{eq:torque_begin}
    \mathbf{\Gamma}_{ji}= \Gamma_0 f(|\mathbf{r}_{ij}|) \,\hat{\mathbf{n}}_i  \times \hat{\mathbf{r}}_{ij},
\end{equation}
which arise in self-aligning active particles \cite{Baconnier2025}, as well as from electrostatic interactions in Janus particles with a metallic (dark) and a dielectric (light) hemisphere \cite{Zhang2021i,Das2024} (\cref{Fig:torques}). The torque $\mathbf\Gamma_{ji}$ exerted by particle $j$ on particle $i$, with amplitude $\Gamma_0$ and a general distance-dependence given by $f(r)>0$ (\cref{Fig:distance_dependence}), tends to turn particle $i$ either towards ($\Gamma_0>0$) or away from ($\Gamma_0<0$) the distance vector $\mathbf{r}_{ij} = \mathbf{r}_{j}-\mathbf{r}_{i} = |\mathbf{r}_{ij}|\, \hat{\mathbf{r}}_{ij}$ connecting it with particle $j$. We define $\Gamma_0$ as the torque amplitude at a distance given by the lattice constant, such that $f(a)=1$. All together, the particles follow the overdamped Langevin equations
\begin{align}
    \frac{\dd}{\dd t} \mathbf{r}_i &= v_0 \hat{\mathbf{n}}_i - \frac{k}{\xi_\mathrm{t}}\Delta\mathbf{r}_i + \sqrt{2D_\text{t}}\,\boldsymbol{\eta}_i^\mathrm{t}, \label{eq:positions}\\
    \frac{\dd}{\dd t} \theta_i &= \frac{1}{\xi_\mathrm{r}}\sum_{j\neq i} \Gamma_{ji} + \sqrt{2D_\text{r}}\,\eta_i^\mathrm{r}, \label{eq:orientations}
\end{align}
where $\xi_\mathrm{t}$ and $\xi_\mathrm{r}$ are the translational and rotational friction coefficients, and $D_\mathrm{t}$ and $D_\mathrm{r}$ are the translational and rotational diffusivities associated with the corresponding Gaussian white noises $\boldsymbol{\eta}_i^{\mathrm{t}}$ and $\eta_i^{\mathrm{r}}$. Here, we indicated the torque as a scalar quantity as it only has a component along the $\hat{\mathbf{z}}$ axis.

Self-propulsion displaces particles away from the lattice sites. Particles reach a displacement $\Delta \mathbf{r}_i = l  \mathbf{n}_i$, with displacement length $l= \xi_\mathrm{t} v_0/k$, in a time scale $\tau_\text{e} = \xi_\mathrm{t}/k$ set by the elastic restoring force (\cref{Fig:lattice}). As in recent work \cite{Hernandez-Lopez2024}, we assume that this elastic relaxation time is much smaller than the time scale of the angle evolution: $\tau_\text{e}\ll \tau_\theta = \xi_\text{r}/\Gamma_0$. Under this approximation, particle positions adiabatically follow the slower orientation dynamics. Ignoring translational noise, which is negligible in front of rotational noise for Janus particles \cite{Zhang2021i,Das2024}, particle positions are given by
\begin{equation} \label{eq:position}
    \mathbf{r}_i(t)  = \mathbf{r}_i^{(0)}+\Delta \mathbf{r}_i (t) = \mathbf{r}_i^{(0)} + l\hat{\mathbf{n}}_i (t),
\end{equation}
where $\mathbf{r}_i^{(0)}$ is the position of the lattice site of particle $i$.

\bigskip

\noindent\textbf{Active crystals as spin lattices}

Under the approximation of fast elastic relaxation, particle positions can be eliminated in favor of the orientations; hence, the active crystal reduces to a spin lattice. To this end, we insert the positions of \cref{eq:position} in \cref{eq:torque_begin} and obtain:
\begin{equation} \label{eq:rescaled_torque}
    \tilde{\mathbf{\Gamma}}_{ji} =\tilde\Gamma_0 \frac{f(|\mathbf{r}_{ij}|)}{|\mathbf{r}_{ij}|} \hat{\mathbf{n}}_i  \times \left[\mathbf{r}_{ij}^{(0)} \edited{+} l\hat{\mathbf{n}}_j\right].
\end{equation}
Here, we made the torque dimensionless by rescaling time as $\tilde t = D_\text{r} t$. As a result, the (signed) dimensionless torque amplitude $\tilde\Gamma_0 \equiv \Gamma_0 /(D_\text{r}\xi_\text{r})$ becomes the only parameter of the system. In \cref{eq:rescaled_torque}, the original polarity-bond interaction $\hat{\mathbf{n}}_i  \times \hat{\mathbf{r}}_{ij}$ between the particles decomposes into two effects: (i) turning either towards or away from the neighbouring lattice site, $\hat{\mathbf{n}}_i\times\mathbf{r}_{ij}^{(0)}$, and (ii) either alignment or anti-alignment with the neighbour's orientation, $\hat{\mathbf{n}}_i\times\hat{\mathbf{n}}_j$.

Assuming nearest-neighbor interactions, and that the displacement $l$ is much smaller than the lattice constant $a$, we expand the radial dependence in powers of $l/a$ as (see \cref{sec:appendix:A})
\begin{equation} \label{eq:distance_expansion}
\frac{f(|\mathbf{r}_{ij}|)}{|\mathbf{r}_{ij}|} \approx \frac{1}{a} \left[ 1 + \left(\Omega - 1 \right) \frac{\mathbf{r}_{ij}^{(0)}\cdot l(\hat{\mathbf{n}}_j - \hat{\mathbf{n}}_i)}{a^2} \right].
\end{equation}
Here, we defined the dimensionless distance-dependence parameter $\Omega\equiv a f' (a)$, which quantifies how the interaction torque depends on distance. It is negative (positive) for torques that decay (grow) with distance (\cref{Fig:distance_dependence}). 
Introducing \cref{eq:distance_expansion} in \cref{eq:rescaled_torque}, we obtain
\begin{multline} \label{eq:torque_spin_vector}
    \tilde{\mathbf{\Gamma}}_{ji}  = \tilde\Gamma_0 \left\{ \left[\frac{1}{a} + \frac{l}{a}\left(\Omega - 1 \right) \frac{\mathbf{r}_{ij}^{(0)}\cdot (\hat{\mathbf{n}}_j-\hat{\mathbf{n}}_i)}{a^2}\right] \hat{\mathbf{n}}_i  \times \mathbf{r}_{ij}^{(0)} \right.\\
    \left. \phantom{\frac{l}{a}\left(\Omega - 1 \right) \frac{\mathbf{r}_{ij}^{(0)}\cdot (\hat{\mathbf{n}}_j-\hat{\mathbf{n}}_i)}{a^2}} + \frac{l}{a} \hat{\mathbf{n}}_i  \times \hat{\mathbf{n}}_j \right\}
\end{multline}
to first order in $l/a$. The first two terms represent the orienting towards neighbouring lattice sites, at the zeroth and first order of the $l/a$ expansion. The third term describes the effective neighbour alignment or antialignment.

\begin{figure}[tb!]
    \centering
  {\phantomsubcaption\label{Fig:XY}}
  {\phantomsubcaption\label{Fig:lattice-alignment}}
  {\phantomsubcaption\label{Fig:mirror}}
    \includegraphics[width=\columnwidth]{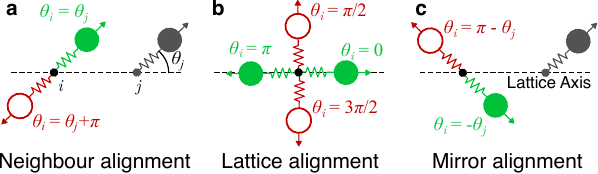}
    \bfcaption{Polarity-bond interactions on a lattice}{ On a lattice (dashed axis), and given a reference particle $j$ (gray), the original interaction torques yield three contributions (\cref{eq:effective_torque}): alignment or anti-alignment with the neighbour's orientation $\propto\sin(\theta_j -\theta_i)$ (\subref*{Fig:XY}), with the lattice axes $\propto\sin 2(\phi_{ij}-\theta_i)$ (\subref*{Fig:lattice-alignment}), and with the neighbour's mirror image $\propto\sin(2\phi_{ij}-\theta_j-\theta_i)$ (\subref*{Fig:mirror}). Green (red) particles indicate the orientations favoured when the prefactor of the corresponding term in \cref{eq:effective_torque} is positive (negative), which depends on the sign of the interaction torques, $\Gamma_0/|\Gamma_0|$, and the distance-dependence parameter $\Omega$.}
    \label{Fig:interactions}
\end{figure}

We now sum over nearest neighbours to obtain the torque on particle $i$:
\begin{multline} \label{eq:effective_torque}
\tilde{\Gamma}_i = \sum_{j\in \langle i,j\rangle} \tilde\Gamma_{ji} = \tilde\Gamma_0 \frac{l}{a} \sum_{j\in \langle i,j\rangle}  \left[ \frac{\Omega +1}{2} \sin(\theta_j -\theta_i) \right. \\
    + \left. \frac{\Omega -1}{2} \left[-\sin 2(\phi_{ij}-\theta_i)  + \sin(2\phi_{ij}-\theta_j-\theta_i) \right]\right].
\end{multline}
Here, the first term of \cref{eq:torque_spin_vector} has cancelled because the lattice vectors $\mathbf{r}_{ij}^{(0)}$ add up to zero for a regular lattice. Thus, all the remaining contributions are of order $l/a$. Hence, this geometric factor sets the magnitude of the torques on a lattice together with the dimensionless torque magnitude $\tilde\Gamma_0$. Moreover, we expressed all the contributions in terms of particle orientations $\theta_i$ and lattice angles $\phi_{ij}$ (\cref{Fig:lattice}, \cref{sec:appendix:B}). The first term in \cref{eq:effective_torque} corresponds to an alignment or anti-alignment torque $\propto \sin(\theta_j -\theta_i)$ like that of the XY model with ferro- or anti-ferromagnetic interactions (\cref{Fig:XY}). The second term produces nematic alignment of a particle $i$ with the lattice axes, given by $\phi_{ij}$, which effectively behave as an external nematic field acting on the spins (\cref{Fig:lattice-alignment}). Finally, the third term produces alignment or anti-alignment of particle $i$ with the mirror image of the neighboring particle $j$; the lattice axis connecting them, encoded in the angle $\phi_{ij}$, acts as the mirror plane (\cref{Fig:mirror}).

The sign of each of these terms depends on the value of $\Omega$, which is determined by the distance dependence $f(r)$ of the interaction torques (\cref{Fig:distance_dependence}). For metal-dielectric Janus colloids \cite{Zhang2021i,Das2024}, their electrostatic dipole-dipole interactions give $f(r) = a^4/r^4$, which gives $\Omega(a)=-4$. For particles reorienting in the chemical concentration field produced by others \cite{Subramaniam2025}, we have $f(r)= a^2/r^2$, which gives $\Omega = -2$. Similarly, systems where torques arise from short-ranged repulsive interactions will have $\Omega <0$. Other possible cases are topological interactions, which are distance-independent, and hence have $\Omega = 0$. Such topological interactions could either be programmed in robots or arise naturally in animals that turn towards or away from their nearest neighbors regardless of their distance. Yet another option is torques due to elastic forces \cite{Ferrante2013}, for which $f(r) = (r-\ell)/(a-\ell)$, and hence $\Omega = 1/(1-\ell/a)$ can be either positive or negative depending on the ratio between the spring's rest length $\ell$ and the lattice constant $a$. Elastic forces were proposed to model the soft interactions between cells \cite{Szabo2006,Henkes2011,Smeets2016}, and they were realised in crystals made of hexbugs connected by springs \cite{Baconnier2022}. Overall, different systems realise different values of the distance-dependence parameter $\Omega$ (\cref{Fig:distance_dependence}). Hence, below we explore its role and we find that it controls the type and strength of orientational order in our active crystals.

Interestingly, the torque in \cref{eq:effective_torque} can be derived from an effective energy $H$, such that the dynamics of the particle orientations $\theta_i$ read
\begin{equation} \label{eq:motion_in_potential}
    \frac{\dd \theta_i}{\dd \tilde t} = - \frac{\partial H}{\partial\theta_i} + \sqrt2\eta^\text{r}_i,
\end{equation}
and
\begin{equation} \label{eq:general_hamiltonian}
    H =  \tilde\Gamma_0 \frac{l}{a} \sum_{\langle i,j \rangle} \left[\frac{\Omega +1}{2} H_{ij}^\mathrm{XY} + \frac{\Omega -1}{2} \left(H_{ij}^\mathrm{LA} + H_{ij}^\mathrm{MA}\right)\right]. 
\end{equation}
This effective energy has contributions due to an XY-type alignment $H_{ij}^\mathrm{XY} = -\cos(\theta_j-\theta_i)$, lattice alignment $H_{ij}^\mathrm{LA} = \left[\cos2(\phi_{ij} - \theta_i) + \cos2(\phi_{ij} - \theta_j)\right]/2$, and mirror alignment $H_{ij}^\mathrm{MA} = -\cos (2\phi_{ij}-\theta_i-\theta_j)$ (\cref{Fig:interactions}). Whereas the distance-dependence parameter $\Omega$ controls the sign and relative strength of these different contributions as discussed above, the turn-towards ($\tilde\Gamma_0>0$) or turn-away ($\tilde\Gamma_0<0$) character of the interaction torques controls the global sign of the effective energy function. Therefore, switching between turn-towards and turn-away torques \cite{Zhang2021i,Das2024} causes a complete inversion of the energy landscape, whereby stable equilibrium points become unstable and viceversa. Such a switch is known as a landscape-inversion phase transition~\cite{Alert2014}, which is of mixed order \cite{Alert2017} and displays unique phase-ordering processes \cite{Alert2016a}.

\bigskip

\noindent\textbf{One-dimensional chain}

\begin{figure}
    \centering
  {\phantomsubcaption\label{Fig:chain}}
  {\phantomsubcaption\label{Fig:phase_diagram}}
  {\phantomsubcaption\label{Fig:order_turn-away}}
  {\phantomsubcaption\label{Fig:order_turn-towards}}
  {\phantomsubcaption\label{Fig:anti-aligned_perpendicular}}
  {\phantomsubcaption\label{Fig:local_anti-ferro_turn-away}}
  {\phantomsubcaption\label{Fig:anti-aligned_along_turn-away}}
  {\phantomsubcaption\label{Fig:anti-aligned_along_turn-towards}}
  {\phantomsubcaption\label{Fig:aligned_perpendicular}}
  {\phantomsubcaption\label{Fig:local_ferro_turn-towards}}
  {\phantomsubcaption\label{Fig:aligned_along}}
    \includegraphics[width=\columnwidth]{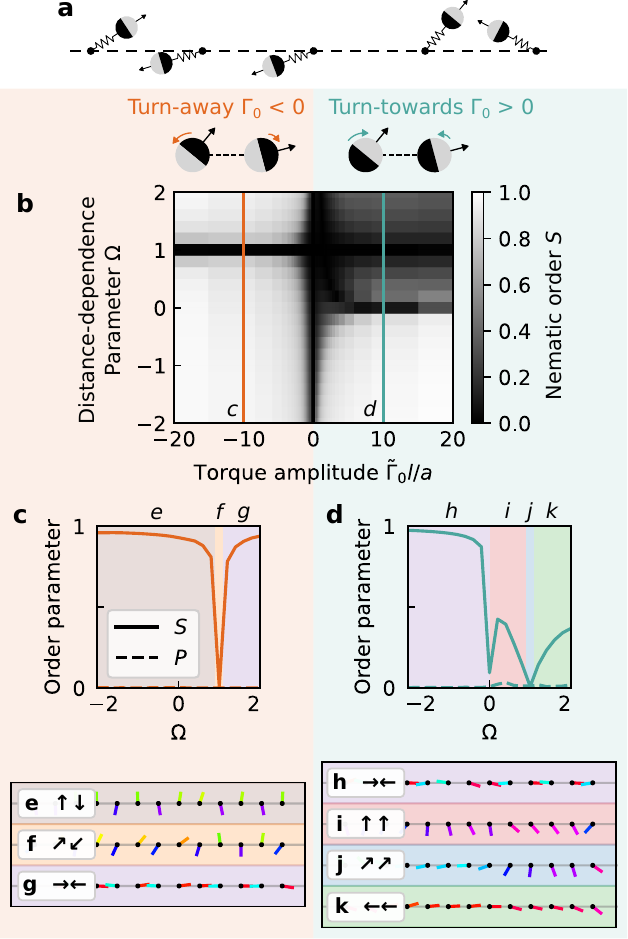}
    \bfcaption{States of particles on a chain}{ \subref*{Fig:chain}, Schematic of a chain of active particles. \subref*{Fig:phase_diagram}, Nematic order $S$ as a function of the dimensionless parameters of the torque interactions: the distance-dependence parameter $\Omega$ and the torque amplitude $\tilde \Gamma_0$. \subref*{Fig:order_turn-away},\subref*{Fig:order_turn-towards}, Polar and nematic order parameters as a function of $\Omega$ for turn-away (\subref*{Fig:order_turn-away}, $\tilde\Gamma_0 l/a = -10$) and turn-towards (\subref*{Fig:order_turn-towards}, $\tilde\Gamma_0 l/a = 10$) torques. Colour shadings indicate different states, shown in snapshots in \subref*{Fig:anti-aligned_perpendicular}-\subref*{Fig:aligned_along}.}
    \label{Fig:chain_simulations}
\end{figure}

To study the emerging orientational order in active crystals with polarity-bond interactions, we start by considering a one-dimensional chain with periodic boundary conditions (\cref{Fig:chain}). We perform Brownian dynamics simulations of \cref{eq:orientations,eq:effective_torque} with $N=10^5$ particles using the Euler method with a time step $d\tilde t = 0.001/(|\tilde \Gamma_0|l/a)$. 
From the simulations, we characterize the emergence of nematic order as a function of the dimensionless parameters $\tilde\Gamma_0$ and $\Omega$ of the torque interactions (\cref{Fig:phase_diagram}).
\edited{In \cref{sec:appendix:C} we show that $N=10^5$ is sufficiently large to avoid finite-size effects, and that the time evolution of the nematic order is not sensitive to different realisations of the random orientations in the initial condition.}

On a chain, each particle has two neighbours with lattice angles $\phi_{ij}=0,\pi$. In this case, the effective energy \cref{eq:general_hamiltonian} reduces to that of an anisotropic XY model for spins $\hat{\mathbf{n}}_i = (\cos\theta_i, \sin\theta_i)$ in a nematic field which aligns them either parallel or perpendicular to the chain axis (see \cref{sec:appendix:D}):
\begin{equation} \label{eq:H_chain}
H = \tilde\Gamma_0 \frac{l}{a} \sum_{i} \left[- \Omega\, n_i^x n_{i+1}^x - n_i^y n_{i+1}^y + \frac{\Omega-1}{2} \cos(2\theta_i) \right].
\end{equation}
Here, the superscripts $x$ and $y$ indicate spatial components. The distance-dependence parameter $\Omega$ controls both the anisotropy of the interactions, reflected in the first two terms, as well as the alignment with the chain axis, encoded in the last term.

Despite the presence of effective alignment interactions, the chain does not exhibit global polar order; the polar order parameter $P =\left\langle \left| \sum_i \hat{\mathbf{n}}_i(t) \right|\right\rangle_t /N$ vanishes (dashed lines in \cref{Fig:order_turn-away,Fig:order_turn-towards}). The situation is reminiscent of the XY model in 1d, for which the Hohenberg-Mermin-Wagner theorem forbids the breaking of the continuous rotation symmetry, and hence the emergence of long-range polar order \cite{Mermin1966,Chaikin1995,Goldenfeld1992}. Here, however, the theorem does not apply because the anisotropy of the interactions as well as the lattice alignment in \cref{eq:H_chain} already break the continuous rotational symmetry.

To rationalise the absence of polar order in our model, we adapt Peierls' argument for the lack of ferromagnetic order in the 1d Ising model \cite{Peierls1936,Goldenfeld1992,Huang1987}. We can extend this argument to our case because the effective energy has the discrete symmetry $\theta_i \to \theta_i + \pi$. Hence, we consider an excitation in the form of a domain of $\pi$-flipped spins, such that the system configuration looks like $\ldots\nearrow\swarrow\ldots\swarrow\nearrow\ldots$ Because of the symmetry of the effective energy function, the domain bulk costs no extra energy; the only energy penalty comes from the domain walls, whose relative contribution decreases with increasing system size $N$. However, the number of ways to place the domain walls, which determines the system's entropy, increases with system size. In the thermodynamic limit $N\to\infty$, and for any non-zero temperature (here noise strength $D_\text{r}>0$), this entropic contribution wins and prevents the emergence of polar order. This argument does not rule out the existence of \emph{local} polar order, as seen in \cref{Fig:aligned_along}. On large scales, however, no polar order persists.

Yet, our active chains are not always disordered. They can display global nematic order (\cref{Fig:phase_diagram}). We quantify it through the scalar nematic order parameter $S=\langle |\sum_{j} e^{i2\theta_j}|\rangle_{t}/N$, which is the largest eigenvalue of the nematic order-parameter tensor $Q_{\alpha\beta}=\langle 2 \,n_{i}^{\alpha}(t)\, n_{i}^{\beta}(t) - \delta_{\alpha\beta} \rangle_{i,t}$, where $\alpha$ and $\beta$ are indices for spatial components. In our system, nematic order arises from the lattice-alignment contribution in the last term of \cref{eq:H_chain}, which acts as an external nematic field with strength controlled by the distance-dependence parameter $\Omega$. For $\Omega = 1$, the lattice-alignment contribution vanishes. In this case, the effective energy \cref{eq:H_chain} corresponds to that of the XY model, for which the Hohenberg-Mermin-Wagner theorem forbids global order. Accordingly, we obtain states with no global nematic order (black horizontal stripe in \cref{Fig:phase_diagram}) but with local order, either ferromagnetic or anti-ferromagnetic (\cref{Fig:local_anti-ferro_turn-away,Fig:local_ferro_turn-towards}).

For other values of $\Omega$, there can be global nematic order (\cref{Fig:phase_diagram}). For small torque amplitudes $\Gamma_0$, fluctuations allow the system to sample different configurations. For large torque amplitudes $\Gamma_0$, the interactions favour specific configurations (\crefrange{Fig:anti-aligned_perpendicular}{Fig:aligned_along}), which we describe and label with arrow symbols below.

For turn-away interactions ($\Gamma_0<0$), we find two states (\cref{Fig:order_turn-away}): anti-aligned perpendicular to the chain (\cref{Fig:anti-aligned_perpendicular}, $\uparrow\downarrow$) and anti-aligned along the chain (\cref{Fig:anti-aligned_along_turn-away}, $\rightarrow\leftarrow$), in addition to the state with only local anti-ferromagnetic order for $\Omega=1$ (\cref{Fig:local_anti-ferro_turn-away}, $\nearrow\swarrow$). Respectively, for turn-towards interactions ($\Gamma_0>0$), we find three states (\cref{Fig:order_turn-towards}): anti-aligned along the chain (\cref{Fig:anti-aligned_along_turn-towards}, $\rightarrow\leftarrow$), aligned perpendicular to the chain (\cref{Fig:aligned_perpendicular}, $\uparrow\uparrow$), and aligned along the chain (\cref{Fig:aligned_along}, $\leftarrow\leftarrow$), in addition to the state with only local ferromagnetic order for $\Omega=1$ (\cref{Fig:local_ferro_turn-towards}, $\nearrow\nearrow$). We note that any of the aligned states described here displays only local polar order. In the following, we explain the emergence of these states by analyzing the equilibrium configurations of two spins.

\bigskip

\noindent\textbf{Two-particle configurations}

\begin{figure} [tb!]
    \centering
    {\phantomsubcaption\label{Fig:two-spin_state_space}}
    {\phantomsubcaption\label{Fig:two-spin_Omega}}
    {\phantomsubcaption\label{Fig:two-spin_states}}
    \includegraphics[width=\columnwidth]{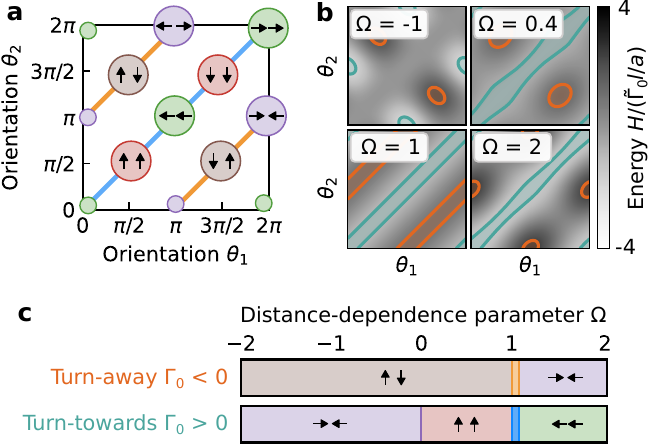}
    \bfcaption{Equilibrium states for two particles}{ \subref*{Fig:two-spin_state_space}, Arrow representations of the two-particle states in the $\theta_1\theta_2$ plane.
    \subref*{Fig:two-spin_Omega}, Effective energy $H$ for different values of $\Omega$. Light (dark) areas are energetically favourable for turn-towards (turn-away) torques, corresponding to $\tilde\Gamma_0>0$ ($\tilde\Gamma_0<0$). The turquoise (orange) contours enclose regions with 90\% of the probability $p(\theta_1,\theta_2)\propto e^{-H}$ for $\Gamma_0=10a/l$ ($\Gamma_0=-10a/l$). \subref*{Fig:two-spin_states}, Most probable two-particle states for $\tilde \Gamma_0 = \pm 10 a/l$, which explain the observations made from the simulations in \cref{Fig:chain_simulations}.}\label{fig:two_particles}
\end{figure}

To understand the states on the chain, we study the equilibrium configurations of two particles on a lattice, described as coupled spins $\theta_1,\theta_2$ governed by the effective energy in \cref{eq:H_chain}. Each point in the $\theta_1,\theta_2$-plane corresponds to a spin configuration as shown in \cref{Fig:two-spin_state_space}. We obtain their effective energy from \cref{eq:H_chain} and show it in \cref{Fig:two-spin_Omega}. For turn-away (turn-towards) interactions, the ground state is given by the minimum (maximum) of $H/\tilde\Gamma_0$.

Beyond the ground state, since the angle dynamics in \cref{eq:motion_in_potential} is equivalent to a system of interacting Brownian particles, the probability density follows the Boltzmann distribution $p(\theta_1,\theta_2) \propto e^{-H(\theta_1,\theta_2)}$. The turquoise (orange) contour lines in \cref{Fig:two-spin_Omega} enclose regions with 90\% of the probability for turn-towards (turn-away) torques at $\tilde \Gamma_0 l/a = 10$ ($\tilde \Gamma_0 l/a=-10$). Using \cref{Fig:two-spin_state_space} as a reference, we identify the states corresponding to these high-probability regions. The results, shown in \cref{Fig:two-spin_states}, match with the states found in our simulations (\cref{Fig:chain_simulations}). Thus, the equilibrium behavior of two particles explains the variety of states found for the many-body system.

Note that, for $\Omega = 1$, the effective energy reduces to that of the XY model, which is rotationally invariant, and hence the minimum becomes degenerate. Accordingly, the probability for turn-towards (turn-away) torques concentrates around the ferromagnetic $\theta_2 = \theta_1$ (anti-ferromagnetic $\theta_2 = \theta_1 +\pi$) ground state, without any preferential alignment with the chain axis (\cref{Fig:two-spin_Omega}, $\Omega = 1$).
For turn-away interactions, the ground states are non-degenerate for $\Omega\neq 1$: They are the $\uparrow\downarrow$ configuration for $\Omega < 1$ and $\rightarrow\leftarrow$ for $\Omega >1$. 
For turn-towards interactions, for $\Omega < 0$, the ground state is also the $\rightarrow\leftarrow$ configuration. For $\Omega \geq 0$, the ground state is degenerate, given by any ferromagnetic configuration $\theta_2 = \theta_1$ (\cref{Fig:two-spin_Omega}, turquoise on the three right-most panels). However, this degeneracy is broken once fluctuations are taken into account, as they allow the particles to explore the shape of the effective energy around the minimum. Analyzing the probability $p(\theta_1,\theta_2) \propto e^{-H(\theta_1,\theta_2)}$ reveals the most likely configurations: $\uparrow\uparrow$ for $0<\Omega <1$, and $\leftarrow\leftarrow$ for $\Omega>1$, as shown in \cref{Fig:two-spin_states}, which match those in \cref{Fig:chain_simulations}.

\bigskip

\noindent\textbf{Lattice-dependent order in two dimensions}

\begin{figure}[tb!]
    \centering
    {\phantomsubcaption\label{Fig:2D_snap}}
    {\phantomsubcaption\label{Fig:2D_closeup}}
    {\phantomsubcaption\label{Fig:3Spin}}
    \includegraphics[width=\columnwidth]{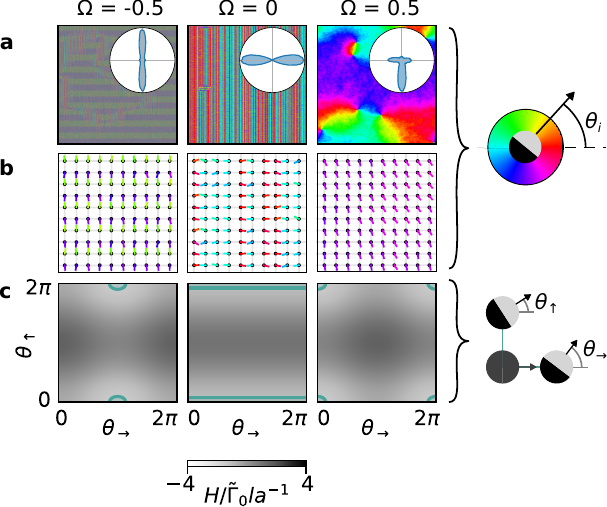}
    \bfcaption{Ordering on a square lattice}{ \subref*{Fig:2D_snap}, \subref*{Fig:2D_closeup}, Snapshots of $500\times 500$ spins (\subref*{Fig:2D_snap}) and close-up views of $10\times 10$ spins (\subref*{Fig:2D_closeup}) for turn-towards interactions with $\tilde \Gamma_0 l/a=50$ and different distance-dependence parameters $\Omega = -0.5,0,0.5$. Color represents particle orientation. The insets in \subref*{Fig:2D_snap} show the angular distribution. \subref*{Fig:3Spin}, Effective energy $H$ between a particle oriented along the $x$ axis and two neighbors: one along the $x$ axis with orientation $\theta_\rightarrow$, and one along the $y$ axis with orientation $\theta_\uparrow$ (see schematic). The turquoise contours enclose regions with 90\% of the probability $p(\theta_\rightarrow,\theta_\uparrow)\propto e^{-H}$.
    } \label{Fig:square_lattice}
\end{figure}

To illustrate that the connection between lattice structure and orientational order extends to two dimensions, we now consider a square \edited{and a triangular} lattice\edited{, respectively with lattice angles $\phi_{ij}=\frac{n\pi}2$ and $\phi_{ij}=\frac{n\pi}3$, where $n=0,1,2,\ldots$.}
\edited{For both lattices,} the lattice-alignment term \mbox{$\propto \sum_{\langle ij\rangle} \cos2(\phi_{ij}-\theta_i)$} of the effective energy in \cref{eq:general_hamiltonian} vanishes, because particles tend to (anti-)align with two perpendicular \edited{or three symmetric} axes. This allows us to study the competition between the neighbour-alignment and mirror-alignment terms in \cref{eq:general_hamiltonian}, tuned by the distance-dependence parameter $\Omega$.

\edited{\textbf{Square lattice} ---} On the square lattice, there is a mapping between any given configuration with turn-away interactions ($\Gamma_0<0$) and another one with turn-towards interactions ($\Gamma_0>0$), as shown in \cref{sec:appendix:E}. Therefore, we focus on the turn-towards case with $\tilde \Gamma_0 l/a=50$ and perform simulations of $500\times 500$ spins with a time step $dt = 0.0005$ for a time $t=5000$ \edited{from an initial condition with random orientations}.

To explore the role of the distance-dependence parameter, we consider the values $\Omega = -0.5,0,0.5$, for which \cref{Fig:2D_snap,Fig:2D_closeup} show snapshots at large and small scales. The amplitude of neighbour and mirror alignment is proportional to $|\Omega +1|$ and $|\Omega-1|$, respectively. For $\Omega =0.5$, neighbor XY alignment is stronger. Consequently, the system forms polar domains and topological defects, similar to the XY model. However, the weak contribution of mirror alignment creates a preference to orient along the lattice, as reflected in the orientational distribution function shown in the inset. For $\Omega = 0$, the neighbor alignment and mirror alignment contributions have equal strengths. In this case, particles orient along one lattice axis, forming a state with nematic order consisting of oppositely-pointing stripes of different widths. For $\Omega = -0.5$, mirror alignment dominates, and the particles are anti-aligned along the direction of orientation and aligned perpendicular to it. This arrangement results in regular stripes with alternating orientation. 

To understand these patterns, we consider particle with a fixed orientation along the $x$ direction ($\theta=0$) and we study the effective interaction energy \cref{eq:general_hamiltonian} for varying orientations of the neighbour in the direction of orientation, $\theta_\rightarrow$, and of the neighbour perpendicular to it, $\theta_\uparrow$. For $\Omega\in(-1,1)$ the neigbour alignment term tends to align both $\theta_\rightarrow$ and $\theta_\uparrow$ with the reference particle, while the mirror alignment term tends to anti-align $\theta_\rightarrow$ and align $\theta_\uparrow$ with $\theta=0$. \Cref{Fig:3Spin} shows the interaction energy $H$, with contour lines enclosing 90\% of the probability $p(\theta_\rightarrow,\theta_\uparrow) \propto e^{-H(\theta_\rightarrow,\theta_\uparrow)}$. For $\Omega = -0.5$, mirror alignment prevails, creating anti-alignment along the orientation direction and alignment perpendicular to it. This is consistent with the aligned stripes of alternating direction seen on large scales (\cref{Fig:2D_snap}, left). For $\Omega = 0.5$, neighbour alignment is stronger, resulting in aligned regions (\cref{Fig:2D_snap}, right). For $\Omega = 0$, both interaction terms tend to align $\theta_\uparrow$ with the reference particle. In contrast, the alignment and misalignment effects on $\theta_\rightarrow$ cancel, such that the interaction does not set the orientation $\theta_\rightarrow$. This is consistent with our observation of stripes that are strongly correlated perpendicular to the particle orientation, but that randomly alternate in the direction of orientation. In all cases, the configurations predicted from this 3-particle picture based on the interaction energy agree with the simulation results.

\begin{figure}[tbh]
    \centering
    {\phantomsubcaption\label{Fig:triangle_XY_away}}
    {\phantomsubcaption\label{Fig:triangle_XY_local}}
    {\phantomsubcaption\label{Fig:triangle_XY_compromise}}
    {\phantomsubcaption\label{Fig:triangle_XY_towards}}
    {\phantomsubcaption\label{Fig:triangle_mirror_away}}
    {\phantomsubcaption\label{Fig:triangle_mirror_local}}
    {\phantomsubcaption\label{Fig:triangle_mirror_compromise}}
    {\phantomsubcaption\label{Fig:triangle_mirror_towards}}
    \includegraphics[width=\columnwidth]{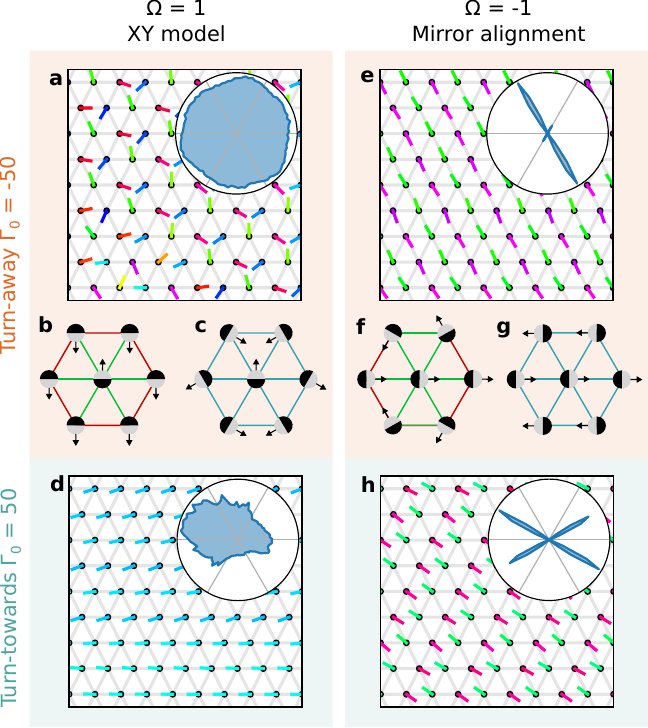}
    \bfcaption{\edited{Ordering and frustration on a triangular lattice}}{ \edited{\subref*{Fig:triangle_XY_away},\subref*{Fig:triangle_XY_towards},\subref*{Fig:triangle_mirror_away},\subref*{Fig:triangle_mirror_towards}, Close-up snapshots of simulations of $500\times 500$ particles on a triangular lattice. Particle orientations are shown in color as in \cref{Fig:square_lattice}. The insets show the distribution of particle orientations. \subref*{Fig:triangle_XY_away},\subref*{Fig:triangle_XY_towards}, The case with $\Omega = 1$ corresponds to the XY model, with either antiferromagnetic ($\Gamma_0<0$) or ferromagnetic ($\Gamma_0>0$) interactions.
    \subref*{Fig:triangle_XY_local},\subref*{Fig:triangle_XY_compromise}, Schematics of frustrated interactions. \subref*{Fig:triangle_XY_local}, For turn-away torques, satisfying the antiferromagnetic interactions of a central particle (green bonds) results in unfavourable interactions between its neighbours (red bonds). \subref*{Fig:triangle_XY_compromise}, The system then reaches a compromise state (turquoise links). \subref*{Fig:triangle_mirror_away},\subref*{Fig:triangle_mirror_towards}, The case with $\Omega = -1$ corresponds to only mirror-alignment interactions, and hence the particles align relative to the lattice. \subref*{Fig:triangle_mirror_local}, For turn-away torques, satisfying the mirror-alignment interactions of a central particle (green bonds) results in unfavourable interactions between neighbours (red bonds). \subref*{Fig:triangle_mirror_compromise}, The system then finds a frustrated compromise state. \subref*{Fig:triangle_mirror_towards}, For $\Omega = -1$, frustration is also present for turn-towards torques.
    }}\label{Fig:frustration}
\end{figure}

\edited{\textbf{Triangular lattice} --- We now consider a triangular lattice and perform simulations for both turn-away and turn-towards torques ($\tilde \Gamma_0 l/a=\pm50$) of $500\times 500$ spins with a time step $dt = 0.0005$ for a time $t=1000$ starting from an initial condition with random orientations (\cref{Fig:frustration}).

We start with the well-known case of the XY model, which we retrieve by setting $\Omega=1$ (see \cref{eq:general_hamiltonian}). In this case, the energy is rotation-invariant and the particles do not align with the lattice, as shown by the orientational distributions in the insets in \cref{Fig:triangle_XY_away,Fig:triangle_XY_towards}. For turn-towards torques ($\Gamma_0 >0$), the particles experience ferromagnetic XY interactions, and hence they develop local polar order (\cref{Fig:triangle_XY_towards}). For turn-away torques ($\Gamma_0<0$), the XY interactions are antiferromagnetic. In a triangular lattice, not all particle pairs can be simultaneously antiparallel (\cref{Fig:triangle_XY_local}), which is known as geometric frustration. As a result, the system reaches states like the one shown in \cref{Fig:triangle_XY_away}, which emerge as a compromise between achieving some anti-alignment between particles while avoiding alignment, as sketched in \cref{Fig:triangle_XY_compromise}.

For $\Omega\neq1$, the effective energy \cref{eq:general_hamiltonian} breaks rotational invariance, and the particles orient relative to the lattice (\cref{Fig:triangle_mirror_away,Fig:triangle_mirror_towards}). To showcase the effects of the lattice, we focus on $\Omega= -1$, for which only the mirror-alignment term in \cref{eq:general_hamiltonian} is present.  
For turn-away torques, \cref{Fig:triangle_mirror_local} shows that satisfying the mirror-alignment interactions for the central particle, again, results in unfavourable interactions between the neighbours. To avoid them, the system reaches a compromise state consisting of alternating aligned stripes, shown in \cref{Fig:triangle_mirror_away} and sketched in \cref{Fig:triangle_mirror_compromise}. For turn-towards torques, the interactions are also frustrated (\cref{Fig:triangle_mirror_towards}). Thus, mirror-alignment interactions are frustrated by the triangular lattice for both signs of the interaction.

For $\Omega\neq\pm1$, an interplay between neighbour and mirror alignment results in generally anisotropic and frustrated states. Particles tend to align relative to the lattice, but due to frustration, the resulting states can no longer be predicted by minimising the interaction energy of two particles as in previous sections.
}

Overall, by extending our analysis of the one-dimensional chain, these results show that lattice-dependent orientational order can also arise in two-dimensional lattices. \edited{In addition, we found that the polarity-bond interactions between our particles can be frustrated in the triangular lattice.} \edited{An interesting question for future work is whether our system can exhibit long-range polar order or not, as discussed for non-reciprocal XY models \cite{Loos2023,Bandini2025,Popli2025,Dopierala2025}.} \edited{Another direction is to consider three-dimensional lattices.}

\bigskip

\noindent\textbf{Discussion and outlook}

In summary, we studied crystals of active particles that turn either towards or away from one another. Because these interactions, which we call polarity-bond interactions, couple the orientation of a particle with the position of another, they establish a link between positional and orientational order. We showed that, when particle positions equilibrate fast compared to their orientations, the orientations can be described as spins that evolve according to an energy. In this energy, the original polarity-bond interactions give rise to both conventional aligning terms like those of the XY model but also unconventional terms that couple the particle orientations to the lattice directions. This energy allowed us to predict the variety of states that we found in direct Brownian dynamics simulations. Thus, our work contributes to ongoing efforts to establish a Hamiltonian description for systems with non-reciprocal interactions which, like our turn-towards or turn-away torques (\cref{eq:torque_begin}), do not obey Newton's law of action and reaction \cite{Shi2025}.

Recent work on active solids showed that the interplay between positional and orientational dynamics gives rise to activity-driven oscillations termed \textit{collective actuation} \cite{Baconnier2022,Xu2023,Baconnier2025}. Here, we explored a different regime by focusing on the limit in which particle positions equilibrate fast compared to their orientations \cite{Hernandez-Lopez2024}. In this regime, our results show that active crystals can display several states with orientational order, with particles aligned in a variety of ways with respect to the lattice directions. The precise state that is favoured depends on whether the interaction torques are turn-towards or turn-away, as well as how they vary with distance.

Thus, our findings reveal that polarity-bond interactions enable one to control the orientational order of active crystals through the lattice structure. Experimentally, such control could be achieved in systems of either metal-dielectric Janus colloids \cite{Yan2016,Zhang2021i,Das2024}, which interact electrostatically through turn-towards or turn-away torques, or macroscopic robots, which can be programmed to do so. Under confinement, active Janus colloids form crystals at high densities due to their repulsive interactions \cite{Das2024}. These repulsive interactions, when approximated for small displacements of the particles around their lattice sites, would give rise to the elastic forces considered in our model. Alternatively, the particles can be placed in engineered lattices made either with grooved substrates \cite{VanBlaaderen1997,Lin2000,Ortiz-Ambriz2016} or with periodic optical potentials generated with interfering lasers \cite{Burns1990,Brunner2002,Mangold2003}. In such lattices, both the structure and the lattice constant can be controlled. In our model, these changes would affect the lattice angles and the value of the distance-dependence parameter $\Omega$, which would then impact the orientational order of the active crystal.

From a theoretical standpoint, our findings introduce the notion of lattice-dependent orientational order, which describes states in which rotational symmetry is broken through a coupling to the lattice structure. By revealing that the lattice structure can impact the orientational order in active crystals, our work complements previous studies of active solids, which mainly focused on how activity distorts or even melts their crystalline structure \cite{Briand2018,Ophaus2021,Shi2023}. \edited{Our work also complements recent studies on the impact of spatial anisotropy, such as the one imposed by a lattice, on flocks \cite{Solon2022,Popli2025,Dopierala2025}.} More generally, our findings call for further developments of general continuum theories of active solids \cite{Maitra2019c,Scheibner2020,Shen2025,Keta2025,Kole2025}: A challenge for future work is to generalise them to incorporate information about the lattice structure which, as we have found, can affect orientational order.

\bigskip

\noindent\textbf{Conflicts of interest}

There are no conflicts of interest to declare.

\bigskip

\noindent\textbf{Data and code availability}

The code produced for this paper is available at \MYhref{https://github.com/tillwelker/Lattice-dependent-orientational-order-in-active-crystals}{this link}.

\bigskip

\noindent\textbf{Acknowledgments}

We thank Mar\'{i}n Bukov for discussions, and for pointing out the mapping of the active chain onto the anisotropic XY model. T.W. thanks Patrick Pietzonka, Holger Stark, and Sarah A.M. Loos for their ideas, inspiration, and support.

\bibliography{Active_crystals}

\onecolumngrid

\clearpage

\twocolumngrid

\appendix

\section{Distance-dependence expansion}\label{sec:appendix:A}

Here, we expand the distance-dependence function $\frac{f(|\mathbf{r}_{ij}|)}{|\mathbf{r}_{ij}|}$ of the torque given in \cref{eq:rescaled_torque} in the Main Text for small displacements of the particles around their lattice sites. The particle position $\mathbf{r}_i = \mathbf{r}_i^{(0)} + l\hat{\mathbf{n}}_i$ given in \cref{eq:position} can be decomposed into the lattice-site position $\mathbf{r}_i^{(0)}$ and the displacement $l\hat{\mathbf{n}}_i$ of the particle from the lattice site. Therefore, the interparticle distance vector reads
\begin{align*}
    \mathbf{r}_{ij} = \mathbf{r}_{j}-\mathbf{r}_{i}= \mathbf{r}_{ij}^{(0)} + l(\hat{\mathbf{n}}_j-\hat{\mathbf{n}}_i) \,,
\end{align*}
with lattice vector $\mathbf{r}_{ij}^{(0)} =  \mathbf{r}_{j}^{(0)}- \mathbf{r}_{i}^{(0)}$ and lattice constant \mbox{$|\mathbf{r}_{ij}^{(0)}|=a$}. If the displacement is much smaller than the lattice constant, $l\ll a$, we can expand the distance-dependence function as 
\begin{align*}
    \frac{f(|\mathbf{r}_{ij}|)}{|\mathbf{r}_{ij}|} \approx  \frac{f(|\mathbf{r}_{ij}|)}{|\mathbf{r}_{ij}|}\Bigg|_{\mathbf{r}_{ij}^{(0)}} + \nabla \frac{f(|\mathbf{r}_{ij}|)}{|\mathbf{r}_{ij}|}\Bigg|_{ \mathbf{r}_{ij}^{(0)}}  \cdot l(\hat{\mathbf{n}}_j-\hat{\mathbf{n}}_i) \,.
\end{align*}
The torque amplitude $\tilde \Gamma_0$ is defined such that $f(a)=1$. Therefore, the zeroth-order term evaluates to $\frac{f(|\mathbf{r}_{ij}|)}{|\mathbf{r}_{ij}|}\Big|_{\mathbf{r}_{ij}^{(0)}} = \frac{1}{a}$. To obtain the first-order term, we calculate the gradient 
\begin{align*}
    \nabla \frac{f(|\mathbf{r}_{ij}|)}{|\mathbf{r}_{ij}|}\Bigg|_{ \mathbf{r}_{ij}^{(0)}} = \left(\frac{\nabla f(|\mathbf{r}_{ij}|)}{|\mathbf{r}_{ij}|} + f(|\mathbf{r}_{ij}|) \nabla \frac{1}{|\mathbf{r}_{ij}|}\right)\Bigg|_{ \mathbf{r}_{ij}^{(0)}} \\
    = \left(\frac{\mathbf{r}_{ij} f'(|\mathbf{r}_{ij}|)}{|\mathbf{r}_{ij}|^2} -\frac{f(|\mathbf{r}_{ij}|)\mathbf{r}_{ij}}{|\mathbf{r}_{ij}|^3}\right)\Bigg|_{ \mathbf{r}_{ij}^{(0)}} = \left(af'(a) -1\right)\frac{\mathbf{r}_{ij}^{(0)}}{a^3}\,.
\end{align*}
Introducing the distance-dependence parameter $\Omega =af'(a)$ and inserting the gradient in the expansion above gives: 
\begin{align*}
    \frac{f(|\mathbf{r}_{ij}|)}{|\mathbf{r}_{ij}|} \approx \frac{1}{a} + (\Omega - 1)\frac{\mathbf{r}_{ij}^{(0)}\cdot l(\hat{\mathbf{n}}_j-\hat{\mathbf{n}}_i)}{a^3}\,.
\end{align*}
This formula is used to obtain the torque between particles on a lattice (\cref{eq:torque_spin_vector} in the Main Text).

\section{Expression of the torque in terms of particle orientations and lattice angles} \label{sec:appendix:B}

Here, we obtain the expression of the torque in  \cref{eq:effective_torque} in the Main Text, which is given in terms of the particle orientations $\theta_i,\theta_j$ and the lattice angles $\phi_{ij}$. We start by summing the interaction torque in \cref{eq:torque_spin_vector} over nearest neighbors:
\begin{align*}
    \sum_{j\in \langle i,j\rangle}\tilde{\mathbf{\Gamma}}_{ji}  = \frac{\tilde \Gamma_0 l}{a}\sum_{j\in  \langle i,j\rangle} \Big[\frac{\hat{\mathbf{n}}_i  \times \mathbf{r}_{ij}^{(0)}}{l}   \hspace{3cm}\\ + \left(\Omega - 1 \right) \frac{\mathbf{r}_{ij}^{(0)}\cdot (\hat{\mathbf{n}}_j-\hat{\mathbf{n}}_i)}{a^2} \hat{\mathbf{n}}_i  \times \mathbf{r}_{ij}^{(0)} 
    + \hat{\mathbf{n}}_i  \times \hat{\mathbf{n}}_j\Big] \,,
\end{align*}
with orientation vector $\hat{\mathbf{n}}_i = (\cos\theta_i,\sin\theta_i,0)$, orientation angle $\theta_i$, lattice vector $\mathbf{r}_{ij}^{(0)} = a (\cos\phi_{ij},\sin\phi_{ij},0)$, and lattice angle $\phi_{ij}$. The first term $\propto \sum_{j\in  \langle i,j\rangle}\hat{\mathbf{n}}_i  \times \mathbf{r}_{ij}^{(0)} = \hat{\mathbf{n}}_i  \times  \sum_{j\in  \langle i,j\rangle}\mathbf{r}_{ij}^{(0)}$ cancels on lattices, where the sum over all lattice vectors vanishes: \mbox{$\sum_{j\in  \langle i,j\rangle}\mathbf{r}_{ij}^{(0)}=0$}.
Note that this sum only vanishes for planar, undeformed lattices like the ones considered here.

Using properties of the dot product and cross products as well as trigonometric identities, we calculate
\begin{align*}
     (\mathbf{r}_{ij}^{(0)} \cdot \hat{\mathbf{n}}_j) (\hat{\mathbf{n}}_i  \times \mathbf{r}_{ij}^{(0)})_z = a^2\cos(\phi_{ij}-\theta_j)\sin(\phi_{ij}-\theta_i) \\
     =\frac{a^2}2 (\sin(2\phi_{ij}-\theta_i-\theta_j)+\sin(\theta_j-\theta_i))\,, \\
    (\mathbf{r}_{ij}^{(0)} \cdot \hat{\mathbf{n}}_i) (\hat{\mathbf{n}}_i  \times \mathbf{r}_{ij}^{(0)})_z = a^2\cos(\phi_{ij}-\theta_i)\sin(\phi_{ij}-\theta_i) \\
    = \frac{a^2}2 \sin2(\phi_{ij}-\theta_i)\,, \\
     (\hat{\mathbf{n}}_i  \times \hat{\mathbf{n}}_j)_z = \sin(\theta_j-\theta_i)\,.
\end{align*}
Inserting these results into the sum of nearest-neighbour torques gives:
\begin{align*}
    \tilde{\Gamma}_{i} = \sum_{j\in  \langle i,j\rangle} \tilde{\Gamma}_{ji} =  \frac{\tilde \Gamma_0 l}{a} \sum_{j\in  \langle i,j\rangle}\Big[ 
    \frac{\Omega-1}{2}\Big( \sin(2\phi_{ij}-\theta_i-\theta_j)\\
    + \sin(\theta_j-\theta_i) -\sin2(\phi_{ij}-\theta_i) \Big) + \sin(\theta_j-\theta_i)\Big] \\
    =   \frac{\tilde \Gamma_0 l}{a} \sum_{j\in  \langle i,j\rangle} \Big[
    \frac{\Omega-1}{2}\Big(\sin(2\phi_{ij}-\theta_i-\theta_j)  \\
    -\sin2(\phi_{ij}-\theta_i) \Big) + \frac{\Omega+1}{2}\sin(\theta_j-\theta_i)\Big]\,.
\end{align*}
This is the expression given in \cref{eq:effective_torque} in the Main Text.

\section{Role of system size and random initial conditions}\label{sec:appendix:C}
\begin{figure}[tbh]
    \centering
    {\phantomsubcaption\label{Fig:finite_away}}
    {\phantomsubcaption\label{Fig:finite_towards}}
    {\phantomsubcaption\label{Fig:noise_away}}
    {\phantomsubcaption\label{Fig:noise_towards}}
    \includegraphics[width=\columnwidth]{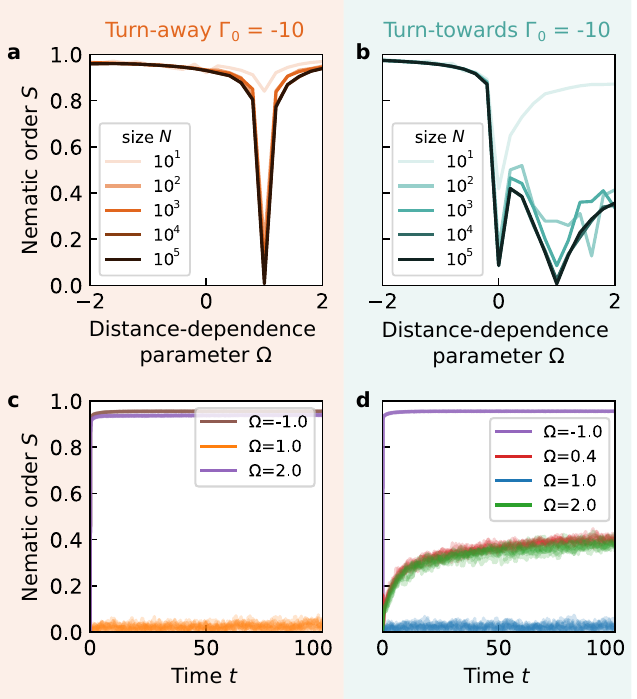}
    \bfcaption{\edited{Role of system size and random initial conditions}}{ \edited{\subref*{Fig:finite_away},\subref*{Fig:finite_towards}, The nematic order $S$ as a function of distance-dependence parameter $\Omega$ for different system sizes $N$ (number of particles) for turn-away (\subref*{Fig:finite_away}) and turn-towards (\subref*{Fig:finite_towards}) torques. \subref*{Fig:noise_away},\subref*{Fig:noise_towards}, Time evolution of the nematic order $S$ for a system of $N=10^4$ particles. For each distance-dependence parameter $\Omega$, 10 different realisations of both the random initial conditions and the noise are shown.}}\label{Fig:finite+noise}
\end{figure}

\edited{In this section, we investigate the role of the system size and the random initial condition for the chain. 

In \cref{Fig:finite_away,Fig:finite_towards} we show the nematic order parameter $S$ as a function of the distance-dependence parameter $\Omega$ for systems with different number of particles $N$. For both turn-away and turn-towards torques, the results become independent of the system size beyond $10^4$ particles. In \cref{Fig:chain_simulations}, we present the results for $N=10^5$.

In \cref{Fig:noise_away,Fig:noise_towards} we show the time evolution of nematic order for $N=10^4$ particles. For each parameter set, we show ten realisations of both the random initial conditions and the noise. In all cases, the evolution is not strongly sensitive to the realisation of initial conditions and noise.
}

\section{Anisotropic XY model in a nematic field}\label{sec:appendix:D}

Here, we derive the expression for the energy of particles on a one-dimensional chain (\cref{eq:H_chain} in the Main Text). On a chain, each particle has two neighbours, with lattice angles $\{\phi_{ij}\}=\{0,\pi\}$ and the set of bonds is $\langle ij\rangle = \{(i,i+1)| i = 1,\dots,N\}$. The energy function in \cref{eq:general_hamiltonian} then takes the form
\begin{align*}
    H = \frac{\tilde \Gamma_0 l}{a}  \sum_{\langle i,j \rangle} \Bigg[\frac{\Omega +1}{2} H_{ij}^\mathrm{XY} + \frac{\Omega -1}{2} \left(H_{ij}^\mathrm{LA} + H_{ij}^\mathrm{MA}\right)\Bigg] \,,
\end{align*}
with
\begin{align*}
    H_{ij}^\mathrm{XY} &= -\cos(\theta_j-\theta_i),\\
    H_{ij}^\mathrm{LA} &= \frac{1}{2}\left[\cos(-2 \theta_i) + \cos(-2\theta_j)\right],\\
    H_{ij}^\mathrm{MA} & = - \cos (\theta_i+\theta_j).
\end{align*}
Rearranging the terms, we have
\begin{align*}
    H = \frac{\tilde \Gamma_0 l}{a}  \sum_{\langle i,j \rangle} \Bigg[H_{ij}^\mathrm{XY} + \frac{\Omega -1}{2} \left(H_{ij}^\mathrm{MA} + H_{ij}^\mathrm{XY}\right)+ \frac{\Omega -1}{2}H_{ij}^\mathrm{LA}\Bigg] \,.
\end{align*}
Given that the orientation vector reads \mbox{$\hat{\mathbf{n}}_i=(\cos\theta_i,\sin\theta_i)$}, we use properties of dot product as well as trigonometric identities to calculate
\begin{align*}
    H_{ij}^\mathrm{XY} + H_{ij}^\mathrm{MA} &= -\cos(\theta_j-\theta_i) - \cos (\theta_i+\theta_j) \nonumber \\
    &= -2\cos(\theta_i)\cos(\theta_j) = -2n_i^xn_j^x \\
    H_{ij}^\mathrm{XY} = -\cos(\theta_j-\theta_i) &= -\hat{\mathbf{n}}_i\cdot \hat{\mathbf{n}}_j  = -n_i^xn_j^x - n_i^yn_j^y \,.
\end{align*}
Inserting these expressions into the energy function gives
\begin{align*}
    H &= \frac{\tilde \Gamma_0 l}{a}   \sum_{\langle i,j \rangle} \Bigg[-n_i^xn_j^x - n_i^yn_j^y - (\Omega -1) n_i^xn_j^x +\frac{\Omega -1}{2}H_{ij}^\mathrm{LA}\Bigg] \\
    &= \frac{\tilde \Gamma_0 l}{a} \sum_{\langle i,j \rangle} \Bigg[-\Omega n_i^xn_j^x - n_i^yn_j^y + \frac{\Omega -1}{2}H_{ij}^\mathrm{LA}\Bigg] \,.
\end{align*}
For the bonds $\langle i,j\rangle$ on a chain, and inserting $H_{ij}^\mathrm{LA}$, the energy can be rewritten as
\begin{align*}
    H = \frac{\tilde \Gamma_0 l}{a} \sum_{i} \Bigg[-\Omega n_i^x n_{i+1}^x-n_i^y n_{i+1}^y+\frac{\Omega-1}{2} \cos(- 2\theta_i) \Bigg]\,,
\end{align*}
which is the energy function of the XY with model where the spins have anisotropic interactions, controlled here by $\Omega$, and are subject to an external nematic field, represented by the last term.

\begin{figure}[tbh]
    \centering
    \includegraphics[width=\columnwidth]{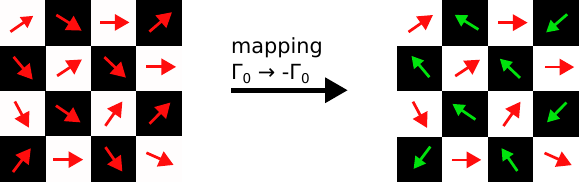}
    \bfcaption{Mapping between turn-away and turn-towards torques on a square lattice}{ Flipping the orientations of particles on all black sites and changing the sign of the interactions $\Gamma_0 \to -\Gamma_0$ leaves the energy invariant.}\label{Fig:2dmap}
\end{figure}

\section{Mapping between turn-away and turn-towards cases on a square lattice}\label{sec:appendix:E}

Here, we show that in a square lattice, for each particle configuration for a system with (signed) torque amplitude $\Gamma_0$, there exists a different particle configuration that has the same energy with torque amplitude $-\Gamma_0$.

On a square lattice, each particle has 4 neighbours with lattice angles $\phi_{ij}=0,\frac\pi2,\pi,\frac{3\pi}2$. The general energy given in \cref{eq:general_hamiltonian} in the Main Text then reads
\begin{align*} 
    H =  \frac{\tilde \Gamma_0 l}{a} \sum_{\langle i,j \rangle} \Bigg[\frac{\Omega +1}{2} H_{ij}^\mathrm{XY} + \frac{\Omega -1}{2} \left(H_{ij}^\mathrm{LA} + H_{ij}^\mathrm{MA} \right)\Bigg] \,,
\end{align*}
with neighbour alignment contribution $H_{ij}^\mathrm{XY} = -\cos(\theta_j-\theta_i)$, lattice alignment contribution $H_{ij}^\mathrm{LA} = \left[\cos2(\phi_{ij} - \theta_i) + \cos2(\phi_{ij} - \theta_j)\right]/2$, and mirror alignment contribution $H_{ij}^\mathrm{MA} = -\cos (2\phi_{ij}-\theta_i-\theta_j)$. The lattice alignment term $H_{ij}^\mathrm{LA}$ vanishes when summing over all lattice angles. If one places the lattice on a checkerboard pattern, particles on white sites interact only with particles on black sites and vice versa, as shown in \cref{Fig:2dmap}. If one flips the orientation of all the particles on black sites, one partner of each interaction is flipped as $\theta_j \to \theta_j + \pi$. This changes the sign of both $H_{ij}^\mathrm{XY}$ and $H_{ij}^\mathrm{MA}$. If one now changes the sign of the interactions by switching between turn-away and turn-towards torques as $\Gamma_0 \to -\Gamma_0$, then the energy remains invariant. The mapping is illustrated in \cref{Fig:2dmap}.

\end{document}